# Structural Resilience of Space-Time Wave Packets to Volumetric Scattering

KEFU MU,[1] HSIAO-CHIH HUANG,[1,2,3], CHEN-TING LIAO,[2,3] AND HUI MIN LEUNG[1,2,3*]

[1]*Department of Intelligence Systems Engineering, Indiana University Bloomington, 700 N Woodlawn Ave, Indiana 87804, USA*
[2]*Department of Physics, Indiana University Bloomington, 727 E. 3rd Street, Bloomington, Indiana 47405, USA*
[3]*Quantum Science and Engineering Center, Indiana University Bloomington, Indiana 47405, USA*
**huileung@iu.edu*

**Abstract:** Optical space–time wave packets (STWPs) can be engineered to sustain tight transverse confinement over long distances. However, whether the requisite precise spatiotemporal correlations underlying this behavior survives volumetric scattering remains unclear. Here, we experimentally demonstrate that STWP light sheets with transverse thicknesses of 11 and 22 µm largely retain their prescribed spatiotemporal structure after transmission through 10 mm thick scattering phantoms ($\mu_s$ = 0.84 mm$^{-1}$). Moreover, as compared to size-matched Gaussian beams, STWPs exhibit greater preservation of their initial spectral intensity distribution across the prescribed space–time domain following scattering. Additionally, the transmitted STWPs remain tightly confined and preserve on-axis intensity decay rates similar to their unscattered counterparts. As the scattering conditions span biologically relevant regimes, the structural and propagation resilience of STWPs to scattering highlights their potential for extended-depth 3D biomedical microscopy.

## 1. Introduction

High-resolution optical microscopy generally requires high-numerical-aperture optics to form a tightly focused excitation and detection volume. This improves lateral resolution but reduces the depth of field and confines diffraction-limited imaging or optical sectioning to a relatively narrow axial range of the sample. This presents practical challenges whenever the sample surface is uneven, moving, or precise focal positioning is impractical, as in endomicroscopy. The challenge arises because high resolution and large depth of field are fundamentally at odds in systems governed by Gaussian beam propagation. Such a constraint is especially consequential in 3D microscopy, such as in confocal and light-sheet imaging, where beam spreading and scattering loss curtail the ability to deliver focused light through thick samples, thereby limiting the accessible imaging volume.

A broad range of strategies has been developed to extend the usable depth of field (DOF), and they can generally be grouped into two families. The first relies on focal-range multiplexing, in which optical energy is distributed over an extended axial interval. This can be achieved spectrally, by exploiting chromatic focal shifts [1,2]; modally via wavefront division and coaxially focusing multiple optical modes [3,4]; or spatially, by dividing the pupil into annular zones with different optical powers [5,6]. The second family comprises propagation-engineered fields, in which the transverse angular spectrum, hence the spatial degree of freedom, is structured to sustain a localized intensity profile over an extended propagation distance. Bessel and Airy beams are canonical examples of this approach [7,8].

More recently, this field-engineering framework for extending the focal range has been generalized beyond purely transverse spatial structuring. In this approach, a prescribed nonseparable correlation is imposed between the transverse spatial frequency, $k_x$, and the temporal frequency, $\nu = \frac{\omega}{2\pi}$, by restricting the spatiotemporal spectrum to a narrow, effectively one-dimensional trajectory defined by the intersection of the light cone with a spectral plane of prescribed tilt [9–11]. The spectral plane imposes a linear relationship between the axial wavenumber, $k_z$, and angular frequency, ω. The slope of this relationship sets the group velocity of the wave-packet envelope. At the same time, the intersection of the spectral plane with the light cone maps the $k_z$-ω relationship onto a corresponding correlation between $k_x$ and ω. Consequently, the joint spatiotemporal spectrum cannot be factorized into independent spatial and temporal spectra and is therefore nonseparable [12], leading to new families of structured optical beams [13,14]. In the ideal limit, each pair of transverse spatial frequencies, $\pm k_x$, is associated with a unique ω. Together, these constraints coordinate the axial phase evolution of the constituent plane-wave components, allowing the wave-packet envelope to translate rigidly at the group velocity prescribed by the tilt of the spectral plane. Because $\frac{d\omega}{dk_z}$ is constant along the spectral trajectory, all constituent spectral components satisfy a common axial group-velocity constraint. The wave-packet envelope therefore translates rigidly without the relative axial dephasing that causes diffractive spreading in conventional beams, enabling propagation-invariant behavior over an extended distance. Fields synthesized according to this principle are termed optical space–time wave packets, or STWPs.

STWPs have been shown to retain a localized transverse profile over extended free-space distances [11] and to self-heal after localized opaque obstructions [15]. Their behavior in scattering media, however, remains incompletely understood. Previous studies have primarily examined output-plane speckle resistance, structural fidelity, and phase stability after thin scattering elements [16], rather than propagation through an extended scattering medium. It therefore remains unknown whether prescribed nonseparable correlations between spatial and temporal frequencies—the defining property underlying propagation invariance—would survive distributed volumetric scattering and offer advantages over conventional Gaussian illumination. Here, we investigate STWP propagation through extended, weakly absorbing tissue-mimicking phantoms with calibrated reduced scattering coefficients. We characterize the transmitted spatiotemporal spectrum and transverse field profile and determine whether propagation invariance persists after passage through the scattering volume. The results are benchmarked against Gaussian wave packets matched in transverse width and spectral bandwidth. By comparing these metrics before and after propagation through phantoms with reduced scattering coefficients representative of biological tissues, we assess the resilience of STWPs to volumetric scattering and their potential for extended-depth optical imaging.

## 2. Overview of STWP design principles

The theoretical framework for space–time wave packets has been established previously [10] and is summarized here to define the design implemented in this study. We consider a one-dimensional STWP that is uniform along y axis, localized along x axis, and propagates along z axis. Its constituent plane-wave components satisfy the free-space dispersion relation

$$k_x^2 + k_z^2 = \left(\frac{\omega}{c}\right)^2, \tag{1}$$

where $k_x$ and $k_z$ are the transverse and longitudinal components of the wave vector, respectively, ω is the angular frequency and $c$ is the speed of light in free space. The tilted spectral plane can be expressed as

$$k_z(\omega) = k_0 + \frac{\omega - \omega_0}{v_g} = k_0 + \left(\frac{\omega}{\mathrm{c}} - k_0\right)\tan\phi. \tag{2}$$

Here $\omega_0$ is the carrier angular frequency, $k_0 = \omega_0/c$, and $v_g$ is the group velocity of the wave-packet envelop associated with the prescribed tilted spectral plane. The angle $\phi$ denotes the tilt of the spectral plane measured from the ω/c axis, such that

$$v_g = c \cot \phi. \tag{3}$$

Combining Eqs. (1) and (2) yields the required correlation between transverse spatial frequency and angular frequency,

$$k_x(\omega) = \pm \sqrt{\left(\frac{\omega}{c}\right)^2 - \left[k_0 + \left(\frac{\omega}{\mathrm{c}} - k_0\right) \tan \phi\right]^2}. \tag{4}$$

To obtain the profile of the required phase mask to construct a STWP light sheet, we use the relation between the local phase gradient along x and the transverse spatial frequency,

$$k_x = \frac{\partial \Psi(x, y)}{\partial x}. \tag{5}$$

Following established derivations, we assume that wavelength is mapped linearly onto the spectrally dispersed coordinate along y and that each spectrally resolved component is assigned a phase ramp that varies linearly along x. Under the narrowband and paraxial approximations, for which $k = \omega/c \approx k_0$ holds, Eq. (4) reduces to the following expression for the phase mask used to generate the STWP light sheet.

$$\Psi(x, y) = \left(\frac{x - x_c}{x_s}\right) \sqrt{\left(\frac{2\Delta\lambda k_0^2 (1 - \tan\phi)}{\lambda_0}\right)\left(\frac{y}{y_s}\right)} = \left(\frac{x - x_c}{x_s}\right) \sqrt{\left(\frac{y}{\alpha\, y_s}\right)}. \tag{6}$$

Here, $x$ and $y$ denote the coordinates on the phase profile, $x_c$ is the center of the phase pattern along x, $x_s$ and $y_s$ are scaling factors used to match the analytical phase pattern onto the physical phase mask, $\lambda_0$ is the center wavelength, and $\Delta\lambda$ is the optical bandwidth. The parameter $\alpha$ controls the phase profile and, consequently, the thickness of the resultant STWP light sheet, with a smaller $\alpha$ corresponding to a thinner light sheet.

## 3. Experimental Methodologies

### *3.1 Pulse shaper for light structuring and generation of STWP*

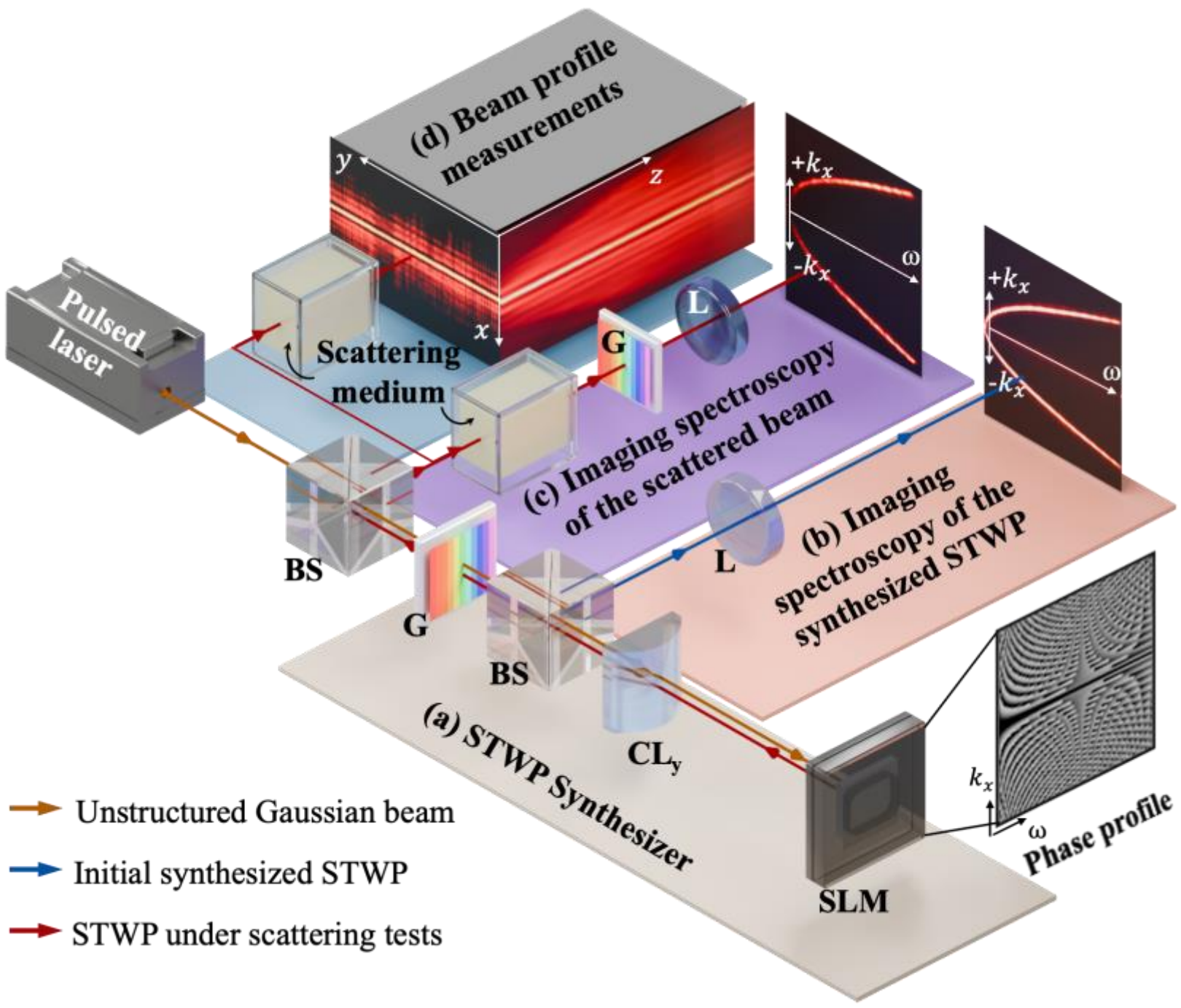

Fig. 1. Experimental setup for STWP synthesis and characterization. (a) The STWP synthesizer is based on a non-rotationally symmetric $4f$ pulse shaper incorporating a programmable SLM. The SLM encodes the phase masks required to generate different STWP configurations, in accordance with Eq. (6). (b) An imaging spectrometer coupled to the pulse shaper is used to measure and verify the $k_x$-ω profile of the synthesized STWP. (c) The $k_x$-ω profiles of the beams, either with or without scattering perturbations, are measured with a custom-built imaging spectrometer. (d) The corresponding evolution of the cross-sectional beam profile in air is captured with a beam-profiling camera translated along the propagation axis. BS: beam splitter, G: grating; $CL_y$: cylindrical lens focusing along the y-axis; SLM: spatial light modulator; L: Lens.

As shown in Fig. 1(a), a non-rotationally symmetric $4f$ pulse shaper comprising a grating, cylindrical lens, and programmable reflective spatial light modulator (SLM) is used to transform a collimated Gaussian pulsed beam into an STWP light sheet, thereby serving as a custom-built STWP synthesizer. A femtosecond laser pulse with a 6 nm bandwidth centered at 1030 nm is first spectrally dispersed along the y-axis by a diffraction grating with a groove density of 1739 lines/mm. The spectrally dispersed field is then collimated along $y$ by a cylindrical lens and directed onto the SLM. The SLM displays the phase profile prescribed by Eq. (6), thereby assigning each resolvable spectral component the requisite transverse spatial frequency $k_x(\omega)$. The modulated field is subsequently reflected back through the pulse shaper, where the spectral components are recombined at the diffraction grating to form the STWP light sheet. This STWP synthesizer forms the basis of the subsequent scattering experiments. To verify the synthesis of the prescribed STWP before the scattering experiments, imaging spectroscopy was used to measure the $k_x$-ω spatiotemporal spectrum of the structured optical field [17]. As shown in Fig. 1(b), this measurement is implemented using a compact modification of a conventional imaging spectrometer configuration. In the conventional arrangement, the field emerging from the pulse shaper is directed to a separate grating–lens–detector assembly, in which the grating resolves the temporal frequencies, ω, along one detector axis and the lens maps the transverse spatial frequencies, $k_x$, onto the orthogonal axis of a 2D detector. Here, the spectral dispersion already produced within the pulse shaper is used directly.

A beam splitter positioned between the pulse-shaper grating and cylindrical lens diverts a portion of the phase-modulated field returning from the SLM, before spectral recombination, toward a Fourier-transform lens and a 2D detector. This setup thus provides the $k_x$-$\omega$ spatiotemporal spectrum without requiring an additional grating.

### *3.2 Determining scattering effects on the beam propagation and spatiotemporal spectrum of STWP with beam profiling and imaging spectroscopy*

We investigate the resilience of STWPs to bulk scattering by determining whether propagation through an extended scattering medium alters their propagation-invariant behavior or spatiotemporal spectrum relative to the unscattered field. The synthesized STWP light sheet is first relayed from the diffraction grating within the pulse shaper to the center of a scattering phantom with a calibrated scattering coefficient, $\mu_s$, as detailed in Section 3.3. After propagating through the 10-mm-thick scattering phantom, the transmitted field is characterized using the optical modules depicted in Figs. 1(c) and (d) to assess the retention of propagation invariance and any changes in its prescribed spatiotemporal spectrum, respectively. To evaluate propagation invariance, the transverse beam profile is recorded over a 50 mm axial range using a beam profiler mounted on an automated translation stage. To characterize the spatiotemporal structure, or equivalently, the $k_x$-ω profile of the transmitted field near the phantom's exit plane, the beam profiler is replaced with a custom-built imaging spectrometer comprising a diffraction grating, a lens, and a 2D detector arranged in a $4f$ configuration [17].

### *3.3 Comparison of STWP scattering resilience with unstructured Gaussian beams*

The Gaussian reference wave packet is generated using the same laser source and optical setup employed for STWP synthesis, as shown in Fig. 1(a). The sole difference is that the hyperbolic STWP phase mask displayed on the SLM is replaced with a spatially uniform, zero-phase pattern, such that no prescribed correlation between spatial and temporal frequencies is introduced. Using the same optical configuration for both the Gaussian and STWP measurements minimizes systematic differences arising from the input beam size and quality, optical path, polarization state, wavefront, and overall beam geometry. To enable a matched comparison of their scattering resilience, the Gaussian reference is focused using a lens of suitable focal length such that the full width at half maximum (FWHM) of its focal spot is comparable to the transverse thickness of the STWP light sheet.

### *3.4 Preparation of phantoms with calibrated effective scattering coefficients*

To mimic the scattering properties of biological specimens, including those of human brain gray matter at 1030 nm, phantoms were prepared using diluted milk solutions. The scattering strength of each phantom was varied by adjusting the volume percentage of milk in water and was calibrated using collimated-transmission measurements acquired with the setup shown in Fig. 2.

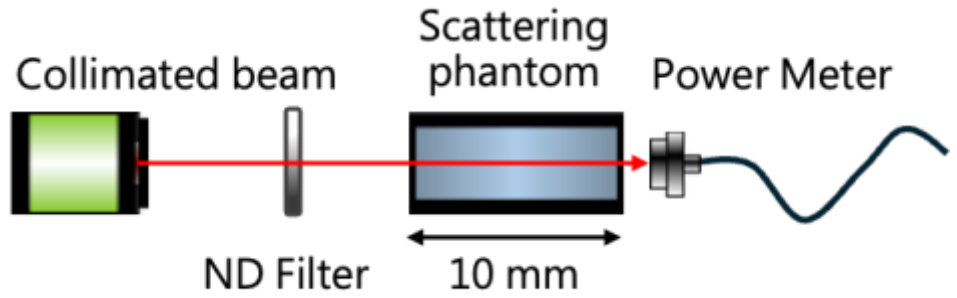

Fig. 2. Setup used for calibrating the scattering coefficients of phantoms. ND: neutral density.

A collimated 1030-nm pulsed laser beam was transmitted through a cuvette ($L = 10$ mm) containing milk solutions of varying concentrations. The transmitted optical power was measured using a power meter positioned close to the exit plane of the cuvette. A background measurement was acquired using the same setup with the laser powered off. The background-subtracted measurements were normalized to a similarly background-subtracted reference transmission acquired without the scattering phantom. Since optical absorption by milk at 1030 nm is substantially weaker than scattering under the present experimental conditions, the measured attenuation was attributed primarily to scattering [18]. The scattering coefficient, $\mu_s$, was therefore estimated using the Beer–Lambert law,

$$I - I_{bg} = \left(I_0 - I_{0,bg}\right) e^{-\mu_s L}$$

where $I$ and $I_0$ are the measured powers with and without the scattering phantom, respectively, $I_{bg}$ and $I_{0,bg}$ are the corresponding background measurements, and $L$ is the propagation distance through the phantom. Using this method, phantoms with scattering coefficients ranging from $\mu_s = 0.22$ to 0.84 mm$^{-1}$ were prepared for the subsequent experiments (Table 1).

**Table 1. Calibrated scattering coefficients of phantoms**

| Concentration (% v/v) | 2 | 5 | 10 | 15 | 20 | 30 | 50 | 100 |
|---|---|---|---|---|---|---|---|---|
| $\mu_s$ (mm$^{-1}$) | 0.22 | 0.33 | 0.55 | 0.76 | 0.84 | 0.88 | 0.97 | 1.02 |

## 4. Results

### *4.1 Establishing the impact of scattering on the spatiotemporal spectrum of the beam*

Using the experimental configuration shown in Fig. 1, we investigated whether STWPs retain their prescribed $k_x$-ω spatiotemporal spectral correlation and propagation-invariant behavior after transmission through a volumetrically scattering medium. Their response was benchmarked against that of conventional Gaussian wave packets matched in beam waist and characteristic $k_x$-ω bandwidth. The Gaussian reference wave packets were generated using the same laser source and pulse-shaping optical train employed for STWP synthesis, as shown in Fig. 1(a). The only difference at the synthesis stage was the phase pattern displayed on the SLM. For the Gaussian reference, the STWP phase mask was replaced with a spatially uniform zero-phase pattern, such that no prescribed correlation between transverse spatial and temporal frequencies was imposed. For each comparison, the Gaussian reference was subsequently focused using a lens of appropriate focal length so that the full width at half maximum of its focal spot follows the transverse thickness of the corresponding STWP light sheet.

Within this size-matched framework, we examined two conditions in which the STWP light sheets had FWHM thicknesses of 22 and 11 $\mu$m and were paired with Gaussian beams having closely matched FWHM beam waists of 24 and 12 $\mu$m, respectively. These paired cases are hereafter referred to as the *coarse* and *fine* confinement conditions. All measurements in this section were performed using the phantom with a scattering coefficient of $\mu_s = 0.84\ \mathrm{mm}^{-1}$, the highest value for which the transmitted signal remained reliably above the detection threshold of the experimental system.

A comparison between the measured STWP $k_x$-$\omega$ spectra before (Fig. 3(b)) and after (Fig. 3(d)) propagation through the scattering phantom indicates that the prescribed spatiotemporal spectral structure remains largely unchanged. In particular, the thickness of the spectral locus, which provides a measure of the uncertainty in the prescribed $k_x$-$\omega$ correlation, as well as the range of $k_x$ values are largely preserved after scattering. Furthermore, scattering led to substantially greater perturbations of the $k_x$-$\omega$ spectrum in Gaussian beams than in STWPs. To quantify the extent of these scattering-induced changes, we calculated the normalized root-mean-square (RMS) difference on a per-pixel basis between the spectra measured after propagation in air and through the scattering medium. The resulting metrics are reported in Table 2. For both the coarse and fine confinement conditions, the normalized RMS difference is lower for the STWP than for the corresponding Gaussian reference, indicating that the STWP more robustly preserves its initial prescribed spectral distribution in $k$-ω space after propagation through the scattering medium as compared to Gaussian beams.

Nevertheless, a portion of the detected spectral intensity is redistributed into a diffuse background concentrated near low $|k_x|$, accompanied by an apparent signal extending to frequencies above the original source bandwidth. However, because the passive scattering medium is not expected to generate new optical frequencies under these experimental conditions, this out-of-band signal is attributed primarily to diffuse scattering of the residual unstructured DC component of the beam and the resultant angular–spectral crosstalk within the imaging spectrometer. The diffuse scattering and, thus, angular broadening of the unstructured DC component can change the range of angles incident on the diffraction grating. This would lead to erroneous mapping of the diffuse light onto detector positions associated with higher apparent frequencies $\omega$. A similar redistribution of detected spectral intensity beyond the original frequency range, concentrated near low $|k_x|$, is also observed for scattered Gaussian beams (e.g., Fig. 3(c)).

**Table 2. Impact of scattering on the spatiotemporal spectrum of Gaussian beams and STWP light sheets**

| Conditions | Coarse | | Fine | |
|---|---|---|---|---|
| | STWP structured beam | Gaussian unstructured beam | STWP structured beam | Gaussian unstructured beam |
| RMS Difference | 0.24 | 0.37 | 0.25 | 0.30 |

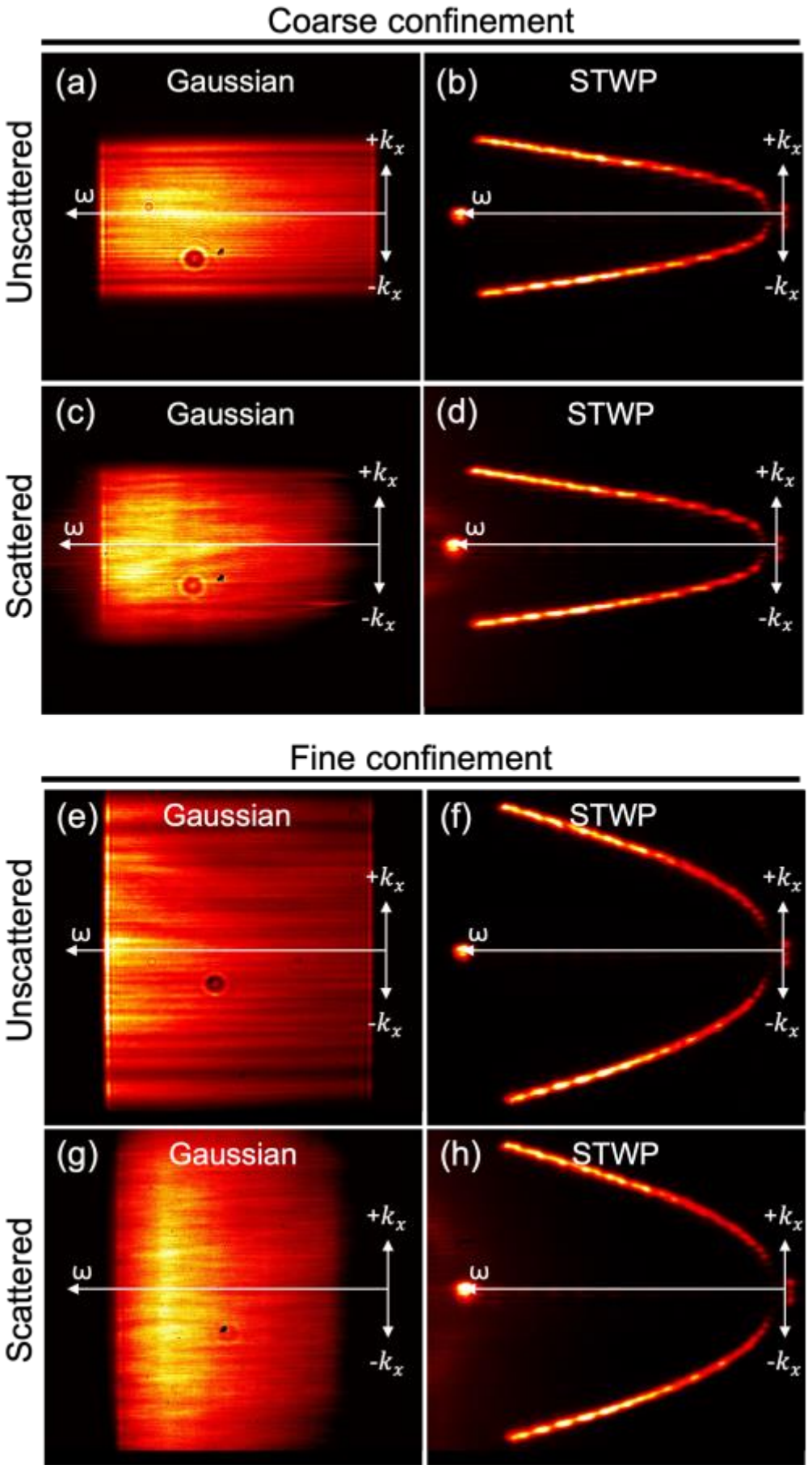

Fig. 3. Comparison of the spatiotemporal spectra of STWP light sheets and Gaussian beams with and without scattering perturbations, under closely size-matched conditions. (a–d) Measurements for the coarse-confinement condition, comprising a 22 µm STWP light sheet and a 24 µm Gaussian beam waist. (e–h) Measurements for the fine-confinement condition, comprising an 11 µm STWP light sheet and a 12 µm Gaussian beam waist.

### *4.2 Establishing the impact of scattering on the transverse confinement of the propagating beam*

We further examined how scattering impacts the transverse confinement of STWPs for both 22 and 11 µm thick light sheets alongside size-matched Gaussian reference beams. Axial–transverse ($x$–$z$) propagation maps were reconstructed by integrating 2D intensity profiles along the $y$-axis over a 50 mm propagation range with 0.5 mm step size downstream of the scattering phantom. Corresponding measurements acquired without the scattering phantom served as the unscattered baseline.

Figures 4(a)–(d) present the free-space propagation maps. They clearly show that the STWP light sheets maintain tight transverse confinement over substantially extended axial distances compared with Gaussian beams, as expected. To quantify this behavior, Fig. 4(e) shows the evolution of the on-axis intensity, defined here as the intensity integrated over a transverse window corresponding to the initial FWHM of each beam and normalized to its value at the propagation origin. The corresponding 3-dB decay distances for the different size-matched cases are summarized in Table 3. The measured 3-dB decay distances of the STWPs are approximately 22-fold greater than those of the unstructured Gaussian beams, demonstrating that STWP structuring confers propagation invariance over extended distances.

Having established this free-space baseline, we next examined propagation following transmission through the volumetrically scattering phantom. For the Gaussian beams, scattering-induced attenuation together with rapid diffractive spreading reduced the detected signal below the detection limit at the phantom exit plane, as illustrated in Figs. 5(b) and 5(d), precluding subsequent tracking of their on-axis intensities. In contrast, the STWPs retained sufficient transverse confinement to remain detectable downstream of the scattering phantom, as illustrated in Figs. 5(a) and 5(c), enabling their post-scattering propagation to be tracked. The corresponding on-axis intensity evolution is shown in Fig. 5(e). Referenced to the phantom exit plane, the 3-dB decay distances in air were 37.5 mm and 19.5 mm for the coarse and fine STWP conditions, respectively. These values are comparable to those measured for the corresponding unscattered STWPs, as summarized in Table 3.

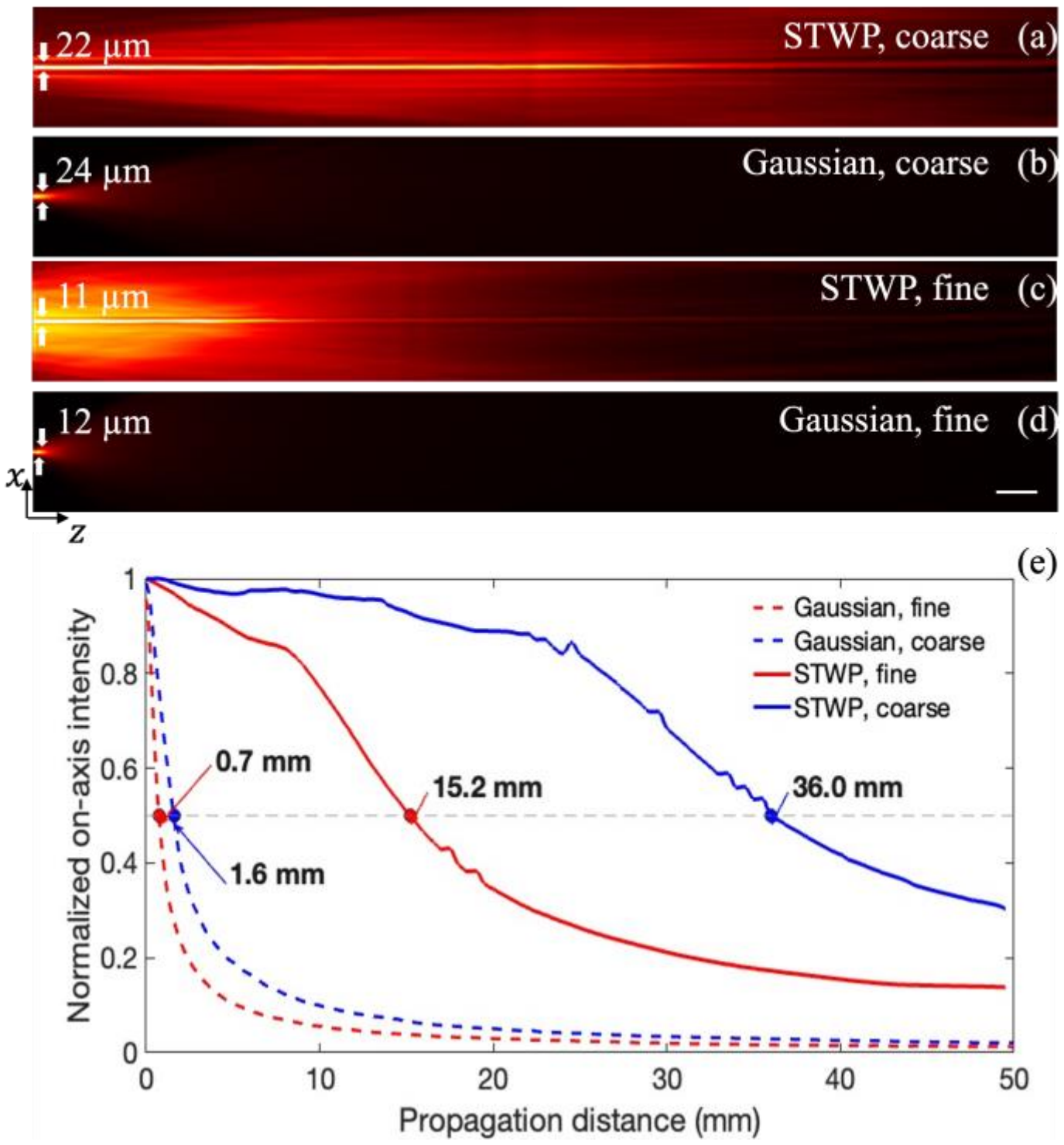

Fig. 4. Baseline characterization of the beam evolution in air, in the absence of scattering perturbations. (a, b) Propagation maps for the coarse-confinement STWP and the corresponding size-matched Gaussian beam. (c, d) Propagation maps for the fine-confinement STWP and its size-matched Gaussian counterpart. Scale bar represents 2 mm. (e) Normalized on-axis intensity as a function of propagation distance for all four beam conditions, with the corresponding 3-dB decay distances indicated.

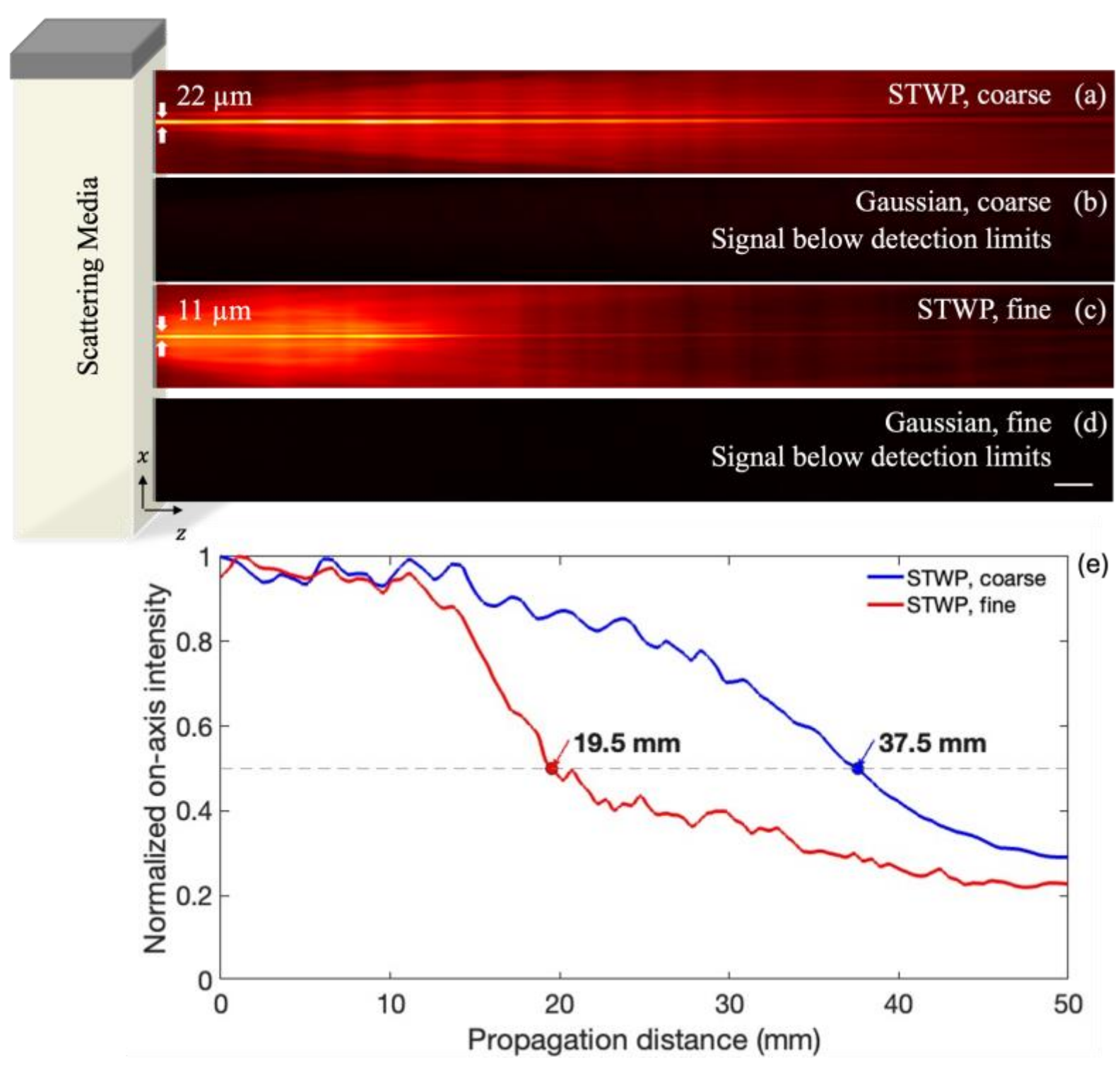

Fig. 5. Characterization of beam evolution in air following transmission through a scattering phantom with $\mu_s = 0.84$ mm$^{-1}$. (a, b) Propagation maps for the coarse-confinement STWP and the corresponding size-matched Gaussian beam. (c, d) Corresponding maps for the fine-confinement condition. In both cases, the Gaussian beams become highly diffuse and fall below the detection limit. Scale bar represents 2 mm. (e) Normalized on-axis intensity as a function of propagation distance, with the 3-dB decay distances indicated where measurable.

**Table 3. On-axis intensity decay of STWP and Gaussian beams with and without scattering perturbations.**

| | **Unscattered** | | | |
|---|---|---|---|---|
| Conditions | Coarse | | Fine | |
| | STWP structured beam | Gaussian unstructured beam | STWP structured beam | Gaussian unstructured beam |
| 3-dB decay distance | 36.0 | 1.6 | 15.2 | 0.7 |
| | **After scattering** | | | |
| 3-dB decay distance from phantom exit plane | 37.5 | Not measurable | 19.5 | Not measurable |

### *4.3 Examination of scattering on the cross-sectional structure of STWP light sheets*

While the preceding analysis focused on the transverse propagation characteristics of the structured and unstructured beams over extended distances, it is also informative to examine how their cross-sectional intensity distributions change under progressively stronger volumetric scattering. Because this evolution cannot readily be captured within the liquid phantom as a function of propagation depth, we instead vary $\mu_s$ while maintaining volumetric scattering over a fixed distance of 10 mm. This approach provides a controlled means of probing the effect of total scattering on the light sheet without assuming strict equivalence to propagation over different depths within a medium of fixed $\mu_s$.

The resulting cross-sectional intensity distributions for the coarse and fine STWPs at the exit plane of the phantom, measured across five scattering phantoms with $\mu_s$ ranging from 0.22 to 0.84 $mm^{-1}$, are shown in Fig. 6(a) to (j). Each cross-sectional intensity map is normalized to its sum intensity to facilitate visualization of changes in the spatial structure of the STWP light sheets across scattering conditions. To assess scattering-induced changes across the transverse beam profile, we first applied a fixed intensity threshold to isolate the light sheet from the background and then evaluated two complementary metrics within the resulting region of interest (ROI). The first was the sum of pixel values within the ROI relative to the reference profile at $\mu_s$=0.22 $mm^{-1}$. The second was the per-pixel RMS difference between each measured light-sheet profile and the same reference, providing a measure of changes in the cross-sectional optical distribution. As shown in Fig. 6(k), the RMS difference increases with $\mu_s$ for both confinement conditions, with a steeper increase for the thinner STWP light sheet. However, this increase closely tracks the corresponding reduction in integrated intensity, indicating that changes to the RMS differences are predominantly driven by scattering-induced attenuation rather than substantial redistribution of optical power within the cross-sectional profile. Assuming $\mu_s L$ approximates the total amounts of scattering, the results shown in Fig. 6(k) is consistent with the faster on-axis intensity attenuation rates for thinner STWP light sheets, as shown in Figs. 4 and 5.

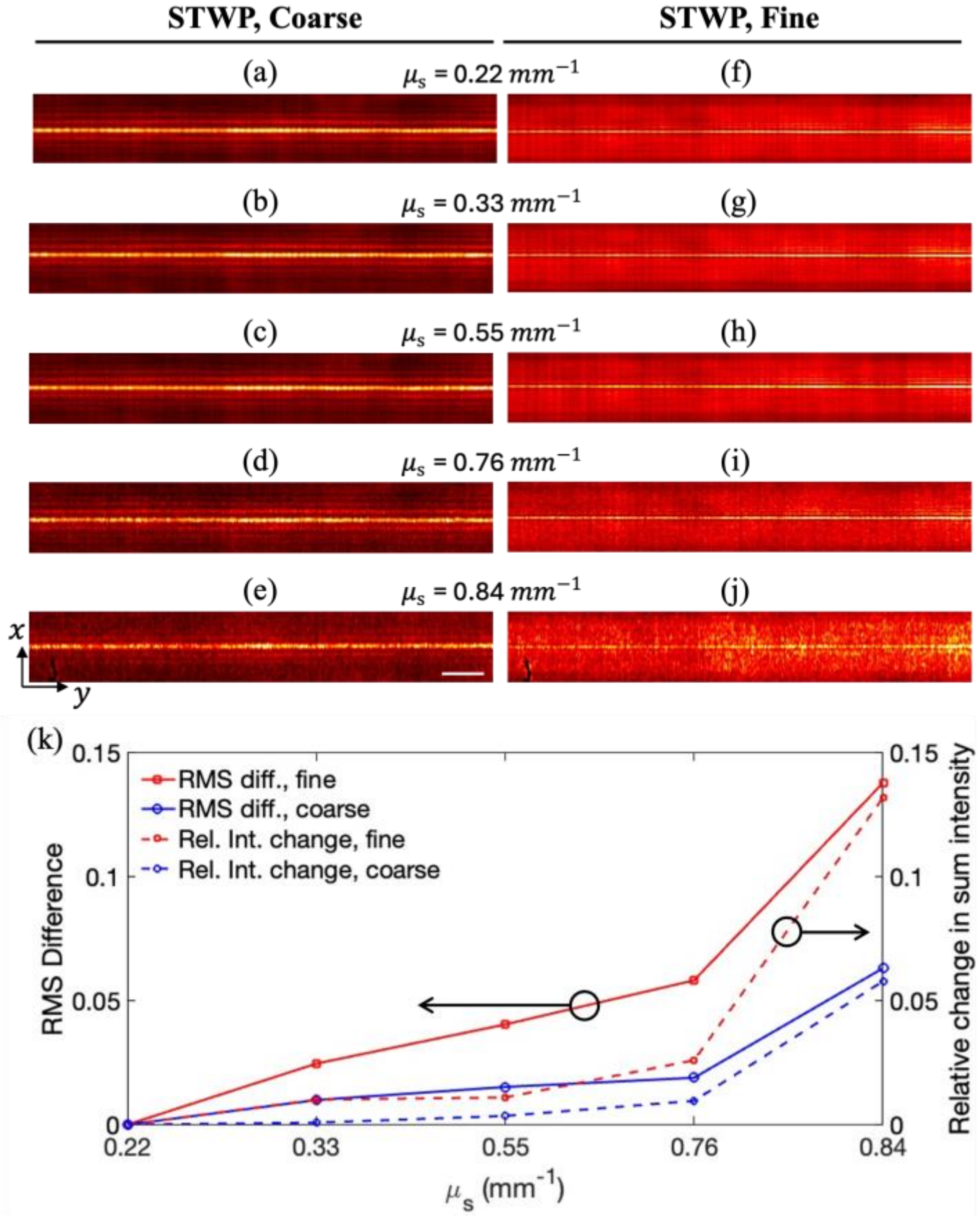

Fig. 6. Cross-sectional beam profiles following transmission through phantoms with different scattering coefficients. (a–e) Measurements for the coarse-confinement STWP light sheet after transmission through phantoms with calibrated scattering coefficients ranging from $\mu_s$ = 0.22 to 0.84 mm$^{-1}$. (f–j) Corresponding measurements for the fine-confinement STWP light sheet under the same scattering conditions. Scale bar represents 200 μm. (k) Solid and dotted lines show the normalized RMS difference (diff.) and the relative change in integrated intensity (Rel. Int. change) within the light-sheet ROI, respectively, as functions of $\mu_s$, with both referenced to the lowest scattering condition, $\mu_s = 0.22$ mm$^{-1}$.

## 5. Discussions

In STWPs, there is a direct importance of engineered space-time structure in conferring tight optical confinement over long distances. Yet it is not fully understood how the spatiotemporal structure is perturbed following propagation through a scattering medium. Our study addresses this gap in knowledge by experimentally examining the extent to which STWPs retain their prescribed spatiotemporal structure and sustain tightly confined light-sheet propagation following transmission through a minimally absorbing scattering medium. The spatiotemporal structure is assessed from the measured $k$-ω spectral profile in the Fourier domain. First, our results show that, although scattering attenuates the optical power of the STWP light sheets in accordance with Beer–Lambert's law, the forward-propagating light emerging from the scattering phantom largely retains the characteristic $k$-ω profile of the STWP, as shown in Fig. 3. Critically, the preservation of this engineered spatiotemporal structure of the laser pulses enables them to continue propagating as tightly confined light sheets downstream of the scattering medium. Furthermore, we examined STWP light sheets with initial FWHM thicknesses of 11 and 22 μm and found that, after transmission through the scattering phantom, not only were their transverse dimensions preserved, their on-axis intensity decay rates also remained comparable to their free-space baselines. These experimental results corroborate the retention of the characteristic spatiotemporal structure of the STWPs after volumetric scattering of the light sheet.

While conventional Gaussian beams are most commonly used for illumination in optical imaging systems, their usable depth of focus is fundamentally constrained by diffraction and the associated Rayleigh range. The engineered spatiotemporal profile of STWPs overcome this limitation to achieve extended depths of focus. In our experiments, the 11 and 22 μm STWP light sheets exhibit 3-dB on-axis decay distances approximately 22 times greater than those of the corresponding size-matched Gaussian beams in free space. More importantly, our results additionally demonstrate that this extended propagation range is preserved following transmission through a volumetrically scattering medium. For the STWPs, the transmitted light that emerged from the bulk scattering medium remains tightly confined to a light sheet similar in thickness as the input, whereas the Gaussian beams undergo rapid diffraction and scattering-induced broadening within the phantom. Consequently, the transmitted field is highly diffuse. These findings suggest that spatiotemporal structuring may provide a valuable strategy for biomedical imaging applications that require sustained tight optical confinement through scattering media. In particular, the ability to preserve a localized forward-propagating field over an extended axial range could improve light delivery to deeper regions of scattering samples and facilitate recovery of spatially localized optical signals for downstream detection.

To place these phantom measurements in the context of biological light delivery, we compared their scattering optical distance with reported near-infrared optical properties of brain gray matter. A scattering coefficient of $5.7\ \mathrm{mm}^{-1}$ has been reported for native human grey brain matter at 1064 nm [19]. Using this reported value and equating the accumulated scattering optical depth as $\mu_s L$, the highest scattering condition investigated here, with $\mu_s = 0.84\ \mathrm{mm}^{-1}$ and $L = 10$ mm, corresponds to approximately 1.47 mm of propagation in gray matter. This distance spans a substantial fraction of the human cortical thickness and reaches the depth of the middle cortical layers, with layer IV located approximately 1.2 to 1.5 mm below the pial surface in representative human cortical tissue. This comparison places the scattering conditions investigated here within an anatomically relevant regime for cortical light delivery, while acknowledging that this comparison strictly serves as an optical depth reference due to experimental simplifications involved in current studies. Although the present experiments examined STWP light sheets with transverse FWHM thicknesses of 11 and 22 μm, the minimum investigated thickness was constrained by the 5.5 μm pixel pitch of the detection camera rather than by the SLM, which is capable of encoding the requisite phase patterns for substantially finer spatial confinement. Therefore, it is possible to extend future work to genuine high-resolution interrogation of biological samples with the use of higher resolution detectors.

Several limitations bound the scope of the present study. Our measurements primarily capture the forward-propagating component that remains above the detection limit after transmission through a minimally-absorbing, uniform, scattering phantom at near-infrared wavelengths. Accordingly, our findings establish the resilience of the engineered spatiotemporal structure within the transmitted field, rather than resilience against scattering attenuation. The extent to which this spatiotemporal structure persists with increasingly tighter confinement, or in heterogeneous scattering media with different anisotropy factors, remains to be determined. Within these bounds, the effects of accumulated scattering on the spatiotemporal structural integrity of the STWPs investigated here support the feasibility of maintaining tightly confined light-sheet delivery across biologically relevant scattering depths.

## 6. Conclusion

In summary, we experimentally established that the underlying spatiotemporal structure enabling STWP light sheets to maintain tight confinement is preserved after volumetric scattering. The scattering-perturbed STWP light sheets retain propagation-invariant behavior over extended distances comparable to that of their unscattered counterparts. These scattering-resilient properties potentially provide STWPs with an advantage over conventional unstructured illumination in applications requiring precise optical access through scattering media, including extended-depth 3D biomedical microscopy, scatterometry, turbidimetry, remote sensing.

**Funding.** This research is supported by the U.S. Department of Energy (DOE) Office of Science, Office of Biological and Environmental Research (BER), grant no. DE-SC0025194.

**Acknowledgment.** The authors gratefully acknowledge Dr. Viet Tran for his assistance in rendering the optical setup schematics used in this work. We also acknowledge the use of generative AI tools (e.g., ChatGPT and Gemini) for language editing and stylistic refinement of the manuscript.

**Disclosures.** The authors declare no conflicts of interest.

**Data availability statement.** Data underlying the results presented in this paper are not publicly available at this time but may be obtained from the authors upon reasonable request.